\documentclass[compsoc,conference,a4paper,10pt]{IEEEtran}
\IEEEoverridecommandlockouts
\usepackage{cite}
\usepackage{amsmath,amssymb,amsfonts}
\usepackage{algorithmic}
\usepackage{graphicx}
\usepackage{textcomp}
\usepackage{bmpsize}
\usepackage{float}
\usepackage{xcolor}
\usepackage{lipsum}
\usepackage{tabularx}
\usepackage{graphicx}
\usepackage{adjustbox}
\usepackage[colorlinks=true,urlcolor=black]{hyperref}
\usepackage{array}
\newcolumntype{L}[1]{>{\raggedright\let\newline\\\arraybackslash\hspace{0pt}}m{#1}}
\newcolumntype{C}[1]{>{\centering\let\newline\\\arraybackslash\hspace{0pt}}m{#1}}
\newcolumntype{R}[1]{>{\raggedleft\let\newline\\\arraybackslash\hspace{0pt}}m{#1}}
\begin{document}

\title{Probing the Mystery of Cryptocurrency Theft:\\ An Investigation into Methods for Taint Analysis
    {\footnotesize
    \thanks{This paper is under review. December 2019.}}}

\author{\IEEEauthorblockN{Tin Tironsakkul}
\IEEEauthorblockA{\textit{School of Mathematical} \\
\textit{and Computer Sciences}\\
\textit{Heriot-Watt University}\\
Edinburgh, United Kingdom \\
tt28@hw.ac.uk}
\and
\IEEEauthorblockN{Manuel Maarek}
\IEEEauthorblockA{\textit{School of Mathematical} \\
\textit{and Computer Sciences}\\
\textit{Heriot-Watt University}\\
Edinburgh, United Kingdom \\
m.maarek@hw.ac.uk}
\and
\IEEEauthorblockN{Andrea Eross}
\IEEEauthorblockA{\textit{School of Social Sciences,} \\
\textit{Accountancy, Economics}\\
\textit{and Finance}\\
\textit{Heriot-Watt University}\\
Edinburgh, United Kingdom \\
a.eross@hw.ac.uk}
\and
\IEEEauthorblockN{Mike Just}
\IEEEauthorblockA{\textit{School of Mathematical} \\
\textit{and Computer Sciences}\\
\textit{Heriot-Watt University}\\
Edinburgh, United Kingdom \\
m.just@hw.ac.uk}
}

% \author{\IEEEauthorblockN{1\textsuperscript{st} Tin Tironsakkul}
% \IEEEauthorblockA{\textit{School of Mathematical and Computer Sciences} \\
% \textit{Heriot-Watt University}\\
% Edinburgh, United Kingdom \\
% tt28@hw.ac.uk}
% \and
% \IEEEauthorblockN{2\textsuperscript{nd} Manuel Maarek}
% \IEEEauthorblockA{\textit{School of Mathematical and Computer Sciences} \\
% \textit{Heriot-Watt University}\\
% Edinburgh, United Kingdom \\
% m.maarek@hw.ac.uk}
% \and
% \IEEEauthorblockN{4\textsuperscript{th} Mike Just}
% \IEEEauthorblockA{\textit{School of Mathematical and Computer Sciences} \\
% \textit{Heriot-Watt University}\\
% Edinburgh, United Kingdom \\
% m.just@hw.ac.uk}
% }

% \IEEEauthorblockN{3\textsuperscript{rd} Andrea Eross}
% \IEEEauthorblockA{\textit{School of Social Sciences,} \\
% \textit{Accountancy, Economics & Finance} \\
% \textit{Heriot-Watt University}\\
% Edinburgh, United Kingdom \\
% a.eross@hw.ac.uk}
% \and

\maketitle
\begin{abstract}
Since the creation of Bitcoin, transaction tracking is one of the prominent means for following the movement of Bitcoins involved in illegal activities. Although every Bitcoin transaction is recorded in the blockchain database, which is transparent for anyone to observe and analyse, Bitcoin's pseudonymity system and transaction obscuring techniques still allow criminals to disguise their transaction trail. While there have been a few attempts to develop tracking methods, there is no accepted evaluation method to measure their accuracy. Therefore, this paper investigates strategies for transaction tracking by introducing two new tainting methods, and proposes an
address profiling approach with a
metrics-based evaluation framework. We use our approach and framework to compare the accuracy of our new tainting methods with the previous tainting techniques, 
%We complete this investigation by applying several tainting methods and our evaluation framework on real theft cases.
using data from two real Bitcoin theft transactions and several related control transactions. 
\end{abstract}


% \textbf{Term Definition} For helping clear up some confusion on what some term really mean, please add comment here if I should put these term in the maint text or clearify it better

% \begin{enumerate}
%   \item \emph{transaction activity}, number of transaction of the address (change from transaction traffic to activity since it seem to make more sense)
%   \item \emph{Service address}, from section 3.3 onward, service address purely means address with very high transaction activity within limited time. So it mean that address is likely to be some point of exchange between individuals. We don't really specify the type of service address in our result yet (explained in section 3.3.1).
%   \item \emph{Transaction Frequency}, for section 3.4.3 actually I am still not sure if I use the right word here. What I want to do is because the thief is going to divide the stolen Bitcoins to a lot of other address, this would make the number of transaction higher than usual, hence the higher transaction frequency? (maybe I can just say transaction amount?)
%   \item \emph{Common transaction/address}, common transaction/address means normal user address and not clean address.
%   \item \emph{Taint analysis}, I change all "tainting analysis" to "taint analysis" to be more in line with other research. But, I will keep using "tainting" when talking about process, method and result 
%   \item \emph{Taint analysis},order-based tainting method is the term I make to identify the FIFO and LIFO tainting that distribute tainted coins based on the transaction order
% \end{enumerate}

\section{Introduction}

Research into cryptocurrencies remains a fascinating topic in many fields due to 
%a
the novel implementation of a decentralised digital currency system. While 10 years have already passed, Bitcoin is still the most prominent and valuable cryptocurrency in use. 
%even though it is no longer the cryptocurrency with the most effective privacy system as of today. 
Bitcoin's high acceptability and pseudonymous privacy capability to protect its users’ identities still makes Bitcoin attractive to individuals who are looking for a less traceable currency, compared to traditional currencies.

%MJ: I edited to make this paragraph two sentences, as it was much too long as a single sentence. 
Even though Bitcoin itself is a decentralised peer-to-peer electronic currency that allows individuals to exchange Bitcoins directly without having to rely on a central third-party entity to monitor or control transactions, the majority of its users 
%continue relying
rely on a controlled third-party entity 
to facilitate the exchange of Bitcoins. These \textit{service entities}, such as cryptocurrency exchange services, help facilitate exchanges to real-world currencies or other cryptocurrencies such as \textit{Coinbase}, or to centralised payment services such as \textit{Bitpay}.

Cryptocurrency service entities are frequently prime targets for individuals that aim to obtain money illegally. Due to the fact that cryptocurrency services such as cryptocurrency exchange services often store their users' Bitcoins in their wallet system to operate the transactions or services, the thefts that occur at cryptocurrency services affect both the service and its users. These events can also undermine the whole economy of cryptocurrencies and affect other cryptocurrencies users. Ultimately, such activities can diminish Bitcoin’s value and its potential to become the official alternative to traditional currencies.

The tracking of Bitcoin transactions remains a formidable challenge due to Bitcoin's privacy protection system and the 
%rising 
rise of new transaction obscuring techniques that allow the individuals engaged in cryptocurrency theft to evade the grasp of law enforcement. While there %are investigations that propose and 
have been proposals for so-called tainting methods that
%develop taint analysis methods in an 
attempt to track the illegal Bitcoins, there 
%is still no investigation 
has been little research
into evaluation criteria to measure the accuracy of tainting results. In light of the above, %mentioned issues, 
this paper contributes the following 
%reusable contributions:
%contributes the following techniques:
\begin{enumerate}
    \item We propose  two new tainting methods, LIFO and TIHO that seek to recognise a thief's strategies for tracking evasion;
    %we introduce to reflect a thieve strategy to evade or hide stolen Bitcoins;
    \item A new approach to tailor tainting with address profiling that only taints transactions up to potential points of theft evasion (e.g., service entities);
    \item A set of metrics to evaluate tainting accuracy that we hypothesise to be indicators of cryptocurrency theft behaviour.
\end{enumerate}
%in this paper 
%we investigate the following research questions:
%\begin{enumerate}
%    \item How can we evaluate the accuracy of the tracking results presented by tainting methods in the case of a given theft transaction?
%    \item In what ways can cryptocurrency tainting methods be improved in order to track the  dissemination of stolen Bitcoins?
%\end{enumerate}
%First, how can we evaluate the accuracy of the tracking results presented by tainting methods in the case of theft transaction?. Second, in what ways can cryptocurrency tainting methods be improved further in order to track dissemination of stolen Bitcoins?.

%Therefore, in this paper, we focus on the development of evaluation metrics for Bitcoin transactions tracking technique (\emph{taint analysis}) by comparing and evaluating tainting strategies. The major contribution of this paper is to explore novel techniques and 
%improve those 
%to evaluate these 
%techniques in order to 
%provide superior 
%improve the
%tracking results for cryptocurrency theft events.

%Ultimately, the aim is to reveal misappropriated cryptocurrencies find their way into the real financial markets.

We apply our approach and evaluation metrics to examine the effectiveness of several tainting methods, including our two new methods. For each of our six evaluation metrics we introduce a corresponding hypothesis that we use to analyse our results. We used data from two real Bitcoin thefts, as well as a number of control transactions for each theft.

%% MJ: The paragraph below doesn't seem to show up well on the pdf for this document style. 
%\paragraph{Plan} 
The rest of the paper proceeds as follows.
In Section~\ref{sec:background}, we give the necessary background on Bitcoin and related work on transaction tracking. We then detail our methodology in Section~\ref{sec:methodology} with the new tainting methods we propose (Section~\ref{new-tainting-methods}), the address profiling we use to tailor tainting (Section~\ref{sec:addressProfiling}), the set of tainting metrics we defined (Section~\ref{sec:tainting-metrics}), and the criteria we used to build the control groups for our experimentation (Section~\ref{control}). We then present in Section~\ref{sec:data} the theft cases we investigate, and discuss the results we obtained in Section~\ref{sec:results}. Section~\ref{sec:conclusion} concludes and presents future work.

\section{Bitcoin Transaction Tainting}
\label{sec:background}

In this section, we provide a brief overview of the Bitcoin system and discuss the tainting methods studied in the previous literature.

% and cryptocurrency theft,

\subsection{Bitcoin System}

% \begin{figure}[H]
% \centering
% {
% \resizebox*{8.2cm}{!}{\includegraphics{Stage.jpg}}}\hspace{5pt}
% \caption{The four stages of cryptocurrency theft. The stage's colour represents the possession of the stolen Bitcoins where the blue colour means the coins are not in the criminal possession and the red colour means the coins are in the criminal possession. In most cases, the cryptocurrency thefts begin with the thief attempting to gain access to the addresses' private key\footnote{A private key is a set of string that act as a password to a Bitcoin address} either by hacking or social engineering. Next, the thief transfers the victim's Bitcoins to their addresses. This stage is typically followed by the attempts to obscure the stolen Bitcoins utilising obscuring techniques. Ultimately, the thief would spend or exchange the stolen coins for real-world value.}
% \label{stage}
% \end{figure}

% The ultimate objective of tracking on illegal activities in cryptocurrencies is to identify the exact transactions that the thief at stage four 

% as shown in Figure~{stage}

The transaction data of Bitcoin is stored in a distributed and transparent transaction ledger system called blockchain, which allows any individual to analyse and visualise every transaction and address in existence~\cite{mcginn_visualizing_2016,giuseppe_bitcoin_2015}. The tracking of the Bitcoin transaction is still not without its challenges, especially in the case of finding the exact ownership and movement of specific Bitcoins.

This is due to the fact that, aside from the pseudonymous address system that helps protect the user's identity and ownership, the possession of Bitcoins in each address exists in the form of unspent transaction outputs (UTXO)\footnote{'Output' is the result of the transaction, which can be used in the subsequent transaction. 'Unspent Transaction Output' (UTXO) is an output that is still unused in any transaction.}, which are newly created from the sum of inputs\footnote{Input is a reference of the previous transaction's output that is being used in the current transaction.} in that transaction. As a result, when the targeted inputs are combined with other unrelated inputs into new output(s), 
%as such this makes it impossible 
it is difficult
to identify or differentiate the exact distribution and destination of the targeted inputs without having a precise methodology.

The primary purpose of the \emph{taint analysis} is to overcome this issue by classifying the targeted Bitcoins (e.g., Bitcoins resulting from a known theft transaction) as tainted (or “dirty”)\footnote{In this paper, we define the tainted Bitcoins as the Bitcoins that contain any portion of the stolen coins, while clean Bitcoins are Bitcoins that do not contain any portion of tainted Bitcoins.}, and any address that uses or transfers them will also be considered as tainted addresses. 
%Next,
Thus,
the tainting method applies a specific rule-set to estimate how the targeted Bitcoins are distributed in the transactions.

The idea of taint analysis is frequently associated with the prospect of regulatory systems implementation on cryptocurrencies, in that addresses identified belonging to criminals would immediately be flagged with a warning system by notifying authorities, business entities and other users that the tainted Bitcoins are in circulation. The tainted Bitcoins should not be accepted by other users or businesses and measures would be taken to place stolen Bitcoins in quarantine; similar to how the blacklisting system works~\cite{bryans_bitcoin_2014,moser_towards_2014}.

\subsection{Taint Analysis Methods}

%The past literature proposes 
We identified
three tainting strategies or methods for tracking transactions using transaction information from the blockchain: 
%which are as follows; 
Poison and Haircut methods~\cite{moser_towards_2014} and FIFO (First In, First Out) method~\cite{ander_fifo_2018}.  %which are discussed in detail in the following sections.
We review these strategies below. 

\subsubsection{The Poison Method} \label{Poison}

The Poison method is a tainting strategy that classifies all of the transaction outputs as tainted outputs in the transactions with any tainted input, regardless of the number of tainted Bitcoins involved \cite{moser_towards_2014}.

\begin{figure}[h]
\centering
{
\resizebox*{2.5cm}{!}{\includegraphics{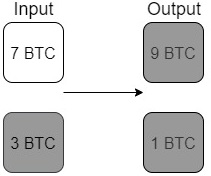}}}\hspace{5pt}
\caption{The Poison Method. The white rectangles represent clean inputs or outputs, while the darker rectangles represent fully tainted ones. For example, in a transaction with a 7 BTC clean input and a 3 BTC tainted input, both of the resulting 9 BTC and 1 BTC outputs will be classified as tainted entirely. The end result is that the number of tainted Bitcoins is now at 10 BTC from initially at 3 BTC.}
\label{poison}
\end{figure}

As shown in Fig~\ref{poison}, the number of the tainted Bitcoins will exponentially increase over time as the tainting continues due to the increase in mixing between the tainted and clean Bitcoins that are in circulation.

The argument for the practicality of this method is that the only way the tainted Bitcoins can affect clean Bitcoins is when they are used together as inputs in the same transaction. Hence, there should be minimal risk of tainted Bitcoins becoming mixed with clean Bitcoins for unrelated users provided that they are cautious not to use the tainted or suspicious Bitcoins they receive in the same transaction with their clean Bitcoins. Additionally, if the users do not use the tainted Bitcoins at all, their addresses will also not be classified as \textit{tainted addresses} (see Section~\ref{sec:addressProfiling}). Therefore, it is possible for criminals to sabotage Bitcoins possession of other unrelated users by purposely sending them a portion of tainted Bitcoins (so that it becomes mixed with other clean Bitcoins)~\cite{moser_towards_2014}.

However, the argument for this method relies heavily on the common user's knowledge of the tainted Bitcoins and strict safety precautions. The Poison method is also frequently considered to be too excessive because of the number of Bitcoins impacted. This makes it impractical to be used for both regulation and tracking purposes compared to other tainting methods. Nevertheless, the Poison tainting results can be used as a baseline method for comparison with other methods.

\subsubsection{The Haircut Method} \label{Hair}

This method operates similarly to the Poison method by also classifying all of the resulting output in the transactions that contained tainted inputs as tainted. However, the Haircut method implements an additional rule: instead of being classified as tainted entirely, each output in the transaction will receive a portion of the tainted inputs according to their proportions~\cite{moser_towards_2014}.

\begin{figure}[h]
\centering
{
\resizebox*{4.5cm}{!}{\includegraphics{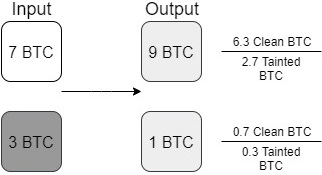}}}\hspace{5pt}
\caption{The Haircut Method. The white rectangles represent clean inputs or outputs, the darker rectangles represent fully tainted ones and the light grey rectangles represent partly tainted ones. Using the same example as the Fig~\ref{poison}, instead of being tainted entirely, both outputs will receive the same proportion of the tainted Bitcoins to the total input value (3/10 proportion), which in this case would be 2.7 tainted BTC for the 9 BTC output and 0.3 tainted BTC for the 1 BTC output.}
\label{haircut}
\end{figure}

As shown in Figure~\ref{haircut}, the Haircut method distributes the inputs based on their proportion in the transaction to each output accordingly. Thus, while both the Poison and Haircut methods consider all of the outputs as tainted in the tainted transactions, the resulting value of the tainted outputs are different. Although, as the mixing between the tainted and clean Bitcoins increases, the tainting result of the Haircut method often accumulates 
%an enormous 
a large
number of tainted transactions and addresses.

Hence, while the Haircut method is more practical than the Poison method  
%for regulation purposes due to a more reasonable tainting methodology, 
the large number of tainted transactions 
%still 
makes 
%this method 
it less suitable for tracking purposes. But similar to the Poison method, the Haircut method can nonetheless provide a useful baseline method of tracking results one can uses as a benchmark to other tainting methods. 
%In this paper, we identify this type of tainting method as \textit{connection-based} tainting method.

\subsubsection{The FIFO Method}

The FIFO (First In, First Out) is a concept of asset inventory management for sorting the order of items via distribution. Hence, we refer to it as \emph{order-based} to distinguish its approach from Poison and Haircut. The concept of FIFO is essentially that the first item that goes in is also the first one that goes out~\cite{ander_fifo_2018}.

\begin{figure}[h]
\centering
{
\resizebox*{4.5cm}{!}{\includegraphics{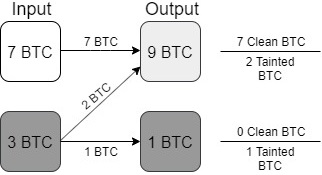}}}\hspace{5pt}
\caption{The FIFO Method. The white rectangles represent clean inputs or outputs, the darker rectangles represent fully tainted ones and the light grey rectangles represent partly tainted ones. With the same transaction example, the FIFO method will start by distributing the first 7 BTC input in the first output. Next, the FIFO method will distribute the 3 BTC input to fill up the rest of the first output. Finally, the remaining of second input will be distributed to the last output. As a result, the first 9 BTC output will contain a portion of the tainted Bitcoins (2 BTC) and the remaining 1 BTC output will contain a full portion of the tainted Bitcoins}
\label{fifo}
\end{figure}

As shown in Figure~\ref{fifo}, the FIFO method determines the distribution order of the transactions: the inputs will be distributed to the outputs from first to last. 
%In this paper, we identify this type of tainting method as \textit{order-based} tainting method.

The argument for implementing the FIFO method as a tainting strategy is that it can provide more precise results compared to the Poison and Haircut methods as FIFO does not consider every resulting output as tainted. This would allow governments or relevant organisations to implement more practical regulations or blacklisting systems. Moreover, the FIFO method is used in the common law (in the United Kingdom) that is created from the historical case in 1816 called Clayton's case for money distribution or withdrawal from bank accounts, thus the use of FIFO method has already been established for official law enforcement in the real world~\cite{ander_fifo_2018}.

However, while the FIFO method possesses valid reasons to be used for regulation purposes, this does not 
%necessary 
necessarily mean it can accomplish the primary objective of cryptocurrency transaction tracking, which is to provide accurate tracking results. This is because the transaction order can be arbitrarily set up in any way possible by the users. The FIFO method that defines the transaction distribution in one specific way 
%would
might be unable to provide accurate tracking results for more complex transactions. It is also possible for the users to circumvent the FIFO method by reverting their transaction order to misdirect the FIFO tracking of their Bitcoins.

\section{Methodology}
\label{sec:methodology}

In this section, we describe our
%the first novel proposed 
two new tainting methods, followed by our address profiling methods that we use in our tainting analysis. We then present our evaluation metrics for measuring tainting performance. For each evaluation metric we present a corresponding hypothesis. 
%Next, we discuss the address profiling methods that we incorporated into our taint analysis. Finally, we discuss the evaluation metrics we develop to evaluate the performance of each tainting method.

\subsection{Proposed New Tainting Methods}
\label{new-tainting-methods}

%The following two tainting methods were motivated by our attempt to consider variations to the original order-based tainting technique, FIFO. In addition to offering two new techniques, a larger set of techniques is useful for studying our evaluation metrics. 

\subsubsection{The LIFO Method}

The LIFO (Last In, First Out) is an alternative method to the FIFO method that operates in the opposite ordering of the FIFO method.
%, and thus it inherits some of the same issues as FIFO, while providing an alternative to correspond to post-theft transaction behaviour. 
LIFO method assumes that the last item that goes in is always the first to goes out.

\begin{figure}[h]
\centering
{
\resizebox*{4.5cm}{!}{\includegraphics{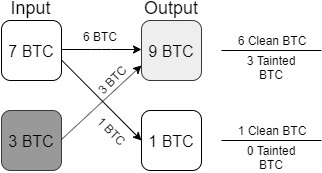}}}\hspace{5pt}
\caption{LIFO method. The white rectangles represent clean inputs or outputs, the darker rectangles represent fully tainted ones and the light grey rectangles represent partly tainted ones. Using the same example as Fig~\ref{poison}, the LIFO method starts by distributing the last 3 BTC input in the first output. Next, the LIFO method distributes the first 7 BTC input to fill up the rest of the first output. Finally, the remaining first input will be distributed to the last output. As a result, the first 9 BTC output will contain a portion of the tainted Bitcoins (3 BTC) and the last 1 BTC output will not contain any tainted Bitcoins.}
\label{lifo}
\end{figure}

As shown in Figure~\ref{lifo}, when applying LIFO method, the last input in the transaction order would be the first to be distributed to the outputs. Next, the LIFO method follows a bottom-up ordering when distributing the inputs.

We implement the LIFO method as a tainting method to evaluate the tracking result of order-based tainting and ascertain whether such tainting techniques can be meaningfully implemented for transaction tracking purpose in the case of cryptocurrency theft transactions.

\subsubsection{The TIHO Method}

%The new order-based tainting method 
TIHO (Taint In, Highest Out)
incorporates a novel tainting classification and the transaction characteristic into the tainting algorithm. This tainting method prioritises the distribution of the tainted inputs to the higher value outputs as shown in Figure~\ref{tiho} below. 
Hence, we refer to this as a \textit{value-based} tainting method. 
%We name this tainting method, \emph{Taint In, Highest Out} or TIHO for short.

\begin{figure}[h]
\centering
{
\resizebox*{4.3cm}{!}{\includegraphics{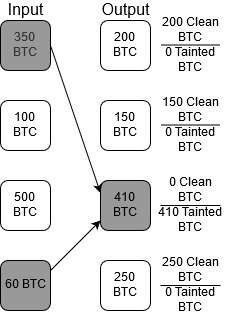}}}\hspace{5pt}
\caption{The TIHO method. The white rectangles represent clean inputs or outputs, the darker rectangles represent fully tainted ones and the light grey rectangles represent partly tainted ones. We use a larger transaction size and transaction value in this example to demonstrate how the TIHO method operate in a more complex transaction. The TIHO method starts by distributing the tainted 350 BTC and 60 BTC inputs to the highest value 410 BTC output. Next, the TIHO method will distribute the remaining clean inputs to the rest of the other lower value outputs. As a result, only the highest value 410 BTC output will be considered as the tainted and the other three outputs will contain no tainted Bitcoins.}
\label{tiho}
\end{figure}

We create this tainting method with the aim to capture larger and more complex transactions that the order-based tainting methods may not be capable to accomplish. TIHO method is developed based on the assumption that the larger value outputs are usually the main purposes of the transaction, while the smaller value outputs are often \textit{change outputs}\footnote{Change output is remaining Bitcoins of the transaction that is being sent back to the senders as other outputs.} of the transaction.

%% MM: The following paragraph is not clear
Although, it should be noted that this assumption does not necessarily apply to every transaction in Bitcoin. For example, in the situation where the users purchase products from merchants using high-value inputs/Bitcoins, similar to using large value banknotes to purchase cheap products, this would mean that the value of the outputs that go to the merchant address would be smaller than the change outputs that go back to the buyers. Hence, the lower value outputs would be the main purpose of the transaction in this example.

\subsection{Address Profiling}
\label{sec:addressProfiling}

Although deanonymisation, which aims to reveal the real identity of Bitcoin addresses' users, is not the goal of this paper, we believe that integrating context-awareness and adaptability capabilities into taint analysis can potentially improve the accuracy of the tracking results. Tainting indiscriminately would miss our goal to understand theft strategies. Therefore, we classify the addresses utilising the information available in the blockchain and tainting results into three categories as follows.

\subsubsection{Service Address} \label{serviceaddress}

As in real life, cryptocurrency services exist in many forms each with different purposes such as cryptocurrency exchange services, online gambling services, E-commerce businesses, marketplace services, CoinSwap services\footnote{CoinSwap is a method of exchange between two or more users through a third party user or a service. The third party would receive the Bitcoins from the sender and send different unrelated Bitcoins to the designated recipients to make the exchange of the Bitcoins untraceable against transaction tracking~\cite{moser2017anonymous}.} and mixing or tumbling services. We define service address as an address that has exceptionally higher transaction activity compared to the other addresses during the same time frame.

We identify service addresses based on the assumption that exceptionally high transaction activity for an address usually indicates that the address is a point of central exchange for other addresses. This is also similar to the concept of degree centrality in network analysis where a high centrality node (a node with many connections to other nodes) often implies that the entity has a large amount of influence to the network and can often be considered as the central hub of exchange for that network~\cite{ghafoor_network_2016}.

We also consider service addresses to be the end goal or exit point of the tainted Bitcoins. We set the assumption that the ultimate purpose of the stolen Bitcoins is to be exchanged for real-world monetary value. As such when tainted Bitcoins reach the service addresses, we can consider the stolen Bitcoins to ultimately reach its uses and are no longer in the hand of individuals engaged in theft. Therefore, we stop the tainting process for any tainted Bitcoins that reach service address. The classification process and the selection criteria for the service addresses are described in more detailed in Section~\ref{sec:results}.

%% MM: The following paragraph is not clear
Although, it should be noted that as we use only the transaction data from the blockchain for our taint analysis, the tainting method can classify the addresses that belong to service entities in the real world as tainted addresses, especially for the services that also avoid reusing their addresses for privacy purposes.

\subsubsection{Tainted Address}

A tainted (or dirty) address is as an address that the tainting methods considers to receive the tainted Bitcoins regardless of the amount. As each tainting method utilises different approaches to tracking, each address may be classified differently in each method. Likewise, it is also possible that the tainted addresses may not belong to the thief, as Bitcoins can also be sent to other users' addresses without having to rely on service entities.

%% MM: The following paragraph is not clear
Thus, the tainted address classification in our experiment is different from the tainting classification proposed in the previous literature in that an address would be considered as tainted when it receive any amount of tainted Bitcoins.

\subsubsection{Clean Address}

A clean address is any address that does not receive any tainted Bitcoin. It should be noted that clean addresses can belong to the theft accomplices depending on how the tainting method operates.

\subsection{Tainting Evaluation Metrics}
\label{sec:tainting-metrics}

We develop evaluation metrics to assess the tainting methods by utilising information available within the blockchain data. Next, we present our hypotheses for each variable. We discuss these evaluation metrics in the subsections below.

\subsubsection{Reused Addresses} \label{privacy measure}

Due to the nature of the transactions involving theft, it is 
%likely 
expected that the thief would employ transaction obscuring and privacy techniques as much as possible to prevent potential tracking. Avoid utilising the same address(es) multiple times, is one of the most common privacy techniques, as the Bitcoin system does not only set any limitation on the number of addresses that the users can possess, but also facilitate the ease of creation of the addresses (in a matter of seconds).

We can assume that the thief would be careful enough to attempt to avoid using an address more than once to reduce transaction traceability. Hence, we consider a reused address is an address that has been used in the transaction more than once. As shown in previous research~\cite{harrigan_unreasonable_2016}\cite{ron_quantitative_2012}, while privacy protection is often considered as one of the most important aspects of Bitcoin among its userbase, many users seem to not be as privacy-conscious as it can be observed from the high number of reused addresses.

This provides us with a valid reason to believe that there is a high chance that the number of reused addresses involved in theft transactions and common transactions is significantly different, giving us the following hypothesis. 
\begin{enumerate}
    \item[H1] The number of reused addresses will be higher in the tainted theft transactions than in the tainted control transactions.
\end{enumerate}
Therefore, 
%we can use this as our hypothesis to evaluate the accuracy of each tainting strategy: 
the sample theft cases should have a considerably lower number of reused addresses compared to the control groups, and that the tainting method that shows the least number of reused addresses is likely to be more accurate.

For our study, reused address metric does not focus on service addresses and considers only the sent transactions because Bitcoin system allows users to send their Bitcoins to any address without requiring confirmation or permission from the receivers as long as they know the public key\footnote{A Public key is an identifier of an address that other users use as a reference for sending Bitcoin.} of the recipient addresses. This means the receiving transactions may not always be in the control of the addresses' owners, while the sending transaction is always in the control of the sender.

\subsubsection{Fresh Addresses}

The fresh address is an address that does not have any transaction activity at all before receiving the tainted Bitcoins. To avoid reusing the same addresses multiple times, the thief would need to create new addresses every time the stolen Bitcoins are distributed. Hence, we hypothesise the following:
\begin{enumerate}
    \item[H2] The tainted theft transactions will have more fresh addresses than those in the tainted control transactions. 
%    \item[H2b] The tainting method with a higher number of fresh addresses will be more accurate.
\end{enumerate}
%Hence, we hypothesise that the sample theft cases would have more fresh addresses than the control groups and that the tainting method that has higher number of fresh addresses is likely to be more accurate.
Thus, we expect that the tainting method with a higher number of fresh addresses will be more accurate.

\subsubsection{Transaction Fee} \label{txfee}

A Transaction fee is an incentive provided by the transaction sender(s) to miner\footnote{A Miner is an individual or a group of individuals that can successfully solve the earliest the block mining challenge provided by the Bitcoin protocol.} to prioritise confirming the transaction into the blockchain. The transaction fee is calculated from the difference between input and output value of the transaction~\cite{nakamoto_bitcoin:_nodate}. Normally, the recommended transaction fee rate that the miners charge is calculated from the data size of the transaction and the number of transactions that are currently waiting for confirmation at that point of time. Although it is possible for the transaction fees to be zero or lower than the recommended rate as the transaction selection for mining is solely determined by the miners' decision. For example, the miner may decide to confirm transactions with zero transaction fees when few transactions are waiting to be confirmed at that time.

We select the transaction fee as one of the evaluation metrics due to its potential to reveal the difference between transactions that involve an illegal activity and common transactions. The transaction fee variable that we use is the transaction fee value and the ratio of the transaction fee to transaction data size like 1 BTC transaction fee value and 0.5 BTC per byte.

\begin{enumerate}
    \item[H3] The amount of the transaction fee in tainted theft transactions will be higher than in tainted control transactions. 
\end{enumerate}

%In this paper, we hypothesise that the amount of the transaction fee in tainted transactions will be higher than normal transactions for 
This hypothesis is motivated by the assumption that the thief will try to obscure his/her transaction trail by rapidly moving the stolen coins; therefore, he/she needs to provide sufficient incentive through the transaction fee to accomplish this. 
As a result, the tainting strategy with better tracking accuracy should have an overall higher average transaction fee for the tainted transactions according to our hypothesis.

\subsubsection{Service Address Reaching}

We anticipate that as the thief would want to spend the stolen coins as soon as possible to minimise the transaction trail - as the longer the stolen coins are still in his/her possession - the higher the chance for it to be detected. 
\begin{enumerate}
    \item[H4] The tainted theft transactions will reach a service address in higher number.
\end{enumerate}
The tainting strategy that shows the higher number of route to any service address is more likely to be more accurate.

\subsubsection{Transaction Frequency}

Another common transaction obscuring or privacy technique is to distribute the Bitcoins in smaller amounts to multiple addresses to increase the difficulty of tracking the original Bitcoins. As such, we expect that the number of tainted transactions per day for the sample theft cases will be higher than the control groups.
\begin{enumerate}
    \item[H5] The number of tainted theft transactions (per day) will be greater than those for the tainted control transactions 
\end{enumerate}
The tainting method providing result with higher transaction frequency is likely to be more accurate.

\subsubsection{Number of Addresses per Transaction}

Similar to the transaction frequency metric, the majority of the transaction-obscuring techniques often involve a large number of addresses in each transaction whether it be laundering services, coin mixing~\cite{ruffing2014coinshuffle,bonneau2014mixcoin}, and CoinJoin\footnote{CoinJoin is a transaction obscuring technique where multiple users share the same transaction whether manually or via a service~\cite{Maurer_2017,Meiklejohn_2015}.} technique.
\begin{enumerate}
    \item[H6] The number of addresses per transaction will be greater for the tainted theft transactions than for the tainted control transactions. 
\end{enumerate}
%Hence, we hypothesise that the number of addresses per transaction in the sample theft case should be different from the control groups and 
The tainting method that shows a higher number of addresses per transaction 
%would likely 
is expected to be more accurate. It should be noted that the addresses per transaction evaluation metric includes both the input and output addresses in the transaction, and also the addresses with clean inputs in the transactions.

\subsection{Control Group Criteria} \label{control}

We select the control groups from 
%every common transactions 
the set of all transactions
that possess the most similar characteristic as the sample theft case, as we define below. 
%so that the tainting result comparison can reveal the difference between the transactions that involve illegal Bitcoins and the common transactions that do not involve illegal Bitcoins. 
In particular, there are three %classifications 
criteria that we 
%devise 
use to select the 
%suitable 
control groups for each sample theft case: 
%as the following;

\begin{enumerate}

\item \textit{Time}. We select the common transactions that occur within the same period as the sample theft cases. In this paper, we set the time criteria to be within one month before and one month after the first time the stolen Bitcoins are being distributed.

\item \textit{Transaction value}. We select 
%the common 
transactions with similar transaction value as the sample theft cases. In this paper, we set the transaction value range criteria to be between $\leqslant$ 1,000 BTC and $\geqslant$ 1,000 BTC, e.g., if the theft is involved in 5,000 Bitcoins, the transaction value criteria for control groups will be set at between 4,000 - 6,000 Bitcoins for that particular theft case.

\item \textit{Transaction distribution}. We select the common transactions that possess a closely similar characteristic as the first transaction after the theft transaction. For example, if the first stolen Bitcoin distribution transaction is a one-to-two addresses transaction, the control groups we pick will also be one address to two addresses transaction.

\end{enumerate}

\section{Data Collection}
\label{sec:data}

%In this paper, 
We use two historical thefts as the samples for testing the tainting methods. The theft cases we 
%are going to 
use are the \textit{Bter theft} from 2015 and \textit{Betcoin theft} from 2012.

Bter is a cryptocurrency exchange service located in China. Its service was shut down in 2017 due to the Chinese government’s ban on the use of cryptocurrencies. The theft occurred on 14 February 2015 at 04:32AM in which the hacker stole 7,170 Bitcoins from the Bter's cold wallet address\footnote{A cold wallet address is a type of address that does not actively connect to any network, or to the internet, to protect against a possible security breach by storing the address file or private key in an offline storage such as USB drive, paper, safe, external hard drive, and an offline computer.}~\cite{stan_bter_2015}. For simplicity, we refer to the Bter theft case sample as "T1" and to their control groups as "C1".

Betcoin or Betco.in is a gambling service that accepts Bitcoin as its primary medium of exchange. The theft occurred on 11 April 2012 at 10:50AM Its service was shut down later after the hack by the service owner~\cite{betcoin_2012,dree12_2012}. As Betcoin requires the users to send their Bitcoins to the service's addresses to be held for the betting, the theft of Betcoin also affects other users, 
%as well 
similar to the thefts of exchange services. For our experiment, we refer to the Betcoin theft case sample as "T2" and to their control groups as "C2".

For this paper, we restricted the taint analysis of the transactions to be within 15 days from the first stolen Bitcoin distribution transaction of each sample case to limit the computational resources and the time required for evaluation. It is also worth noting that while the theft transaction of T2 happened in 2012, the distribution of the stolen Bitcoins began in 2013 which means that the control groups we chose for this case 
%will also be 
were from 2013.

\subsection{Service Addresses Classification}

For each sample theft case, we looked at the number of transactions 
%number of 
for every address within the 15 day tainting period. We 
%limited 
set the service address classification to be the addresses that have the number of transactions at the top 99th percentile of all the addresses in the time limit. Thus, the criteria for service addresses are any address with more than 24 transactions for T1 and 28 transactions for T2.

%% MJ: I find the following two paragraphs a bit confusing. I think that there are a couple of points that highlight remaining limitations of using the 99th percentile, but I'm not entirely sure. 
%The reason why 
We choose the 99th percentile limit in order to reduce the risk of misclassification as the previous research~\cite{ron_quantitative_2012,harrigan_unreasonable_2016} shows that the majority of the addresses normally have less than five transactions throughout their lifetime. This means that the higher criteria limit we select, the less chance the common addresses will be misclassified as service addresses. %Although 
The previous research also points out that there are still many individuals who reuse their addresses, despite the ease of address creation process. Hence, common addresses can still have enough transaction activity to be in the top percentile.

It is also worth mentioning that choosing the lower criteria limit would mean a higher chance to include services that employ transaction obscuring techniques such as laundering service addresses, which are likely to be involved in some way. We further address our plan for this type of situation in Section~\ref{sec:results}.

\subsection{Control Groups Selection}

To select the control groups for each sample theft case, we select the common transactions that fit the control group criteria and that do not appear in the Poison tainting result of the sample cases. As multiple transactions in the same transaction set can fit the criteria (e.g., users sending the same Bitcoins to their addresses in the same way multiple times would make all of the transactions fit the control group criteria), we group the transaction sets together and select only the first transaction of the group to avoid redundant control groups and tainting process. However, this does not eliminate the possibility of transaction merging between the control groups later on in the tainting process.

There are 25 control samples for C1 and 14 control samples for C2 in this experiment. However, we had to slightly adjust the transaction distribution characteristic criteria for T2 as the first distribution transaction has shared inputs with seven other addresses that are directly unrelated to the theft case. As employing this criteria would result in too small number of control groups, we ease the distribution criteria for T2 to consider only the single input address of the theft transaction. As a result, the transaction distribution characteristic criteria for T2 is the transactions with a single address input and a single address output. The completed list of the control groups' transaction hash for each case is displayed in Table~\ref{txhashtable} in the appendix.

\section{Results and Discussion} \label{sec:results}

In this section, we present the results and dicussion of our taint analyses on the sample theft cases and their control groups. First, we present our findings for T1 followed by T2 along with the control groups' results in comparison for both theft cases. In the result figures, we abbreviate the word "transaction" as "TX" and "address" as "ADR". The transaction fee metric is presented in Sat or Satoshis\footnote{Sat or Satoshis is the smallest unit in Bitcoin value. One Bitcoin is equal to 100,000,000 Satoshis.} per byte.

As we incorporate the address profiling into each tainting method, the results would be different from the ones originally proposed in the previous literature. Hence, from this point forward, we indicate the inclusion of the address profiling in the tainting methods with ``\textsuperscript{AP}'' (short for address profiling) or ``(AP)'' after the method name.

As the Poison\textsuperscript{AP} and Haircut\textsuperscript{AP} methods classify every output in the transaction as tainted, albeit with different taint value, the tainted transaction results of both methods would be the same including the addresses involved. Hence, we combine the tainting results of Poison\textsuperscript{AP} method into Haircut\textsuperscript{AP} method in the results.

\subsection{The Transaction Frequency Metric}

\begin{figure}[h]
\centering
{
\resizebox*{7.5cm}{!}{\includegraphics{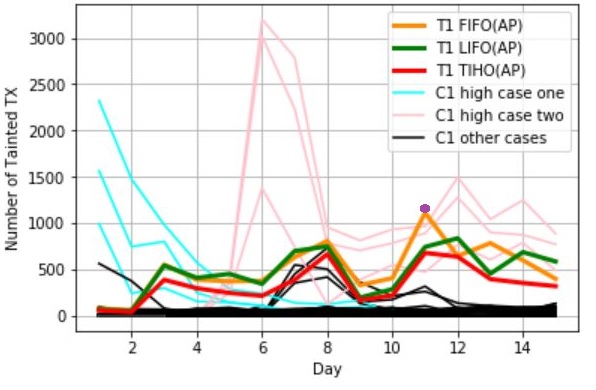}}}\hspace{5pt}
\caption{The transaction frequency per day in T1 and C1 tainting results. Each black line represents the results of each tainting method for each C1 case, and the cyan and pink line represent the two C1 cases with the highest results which are referred as C1 high case one and two. The orange, green and red lines represent each tainting method results in T1. The purple dot at day 11 on T1 FIFO\textsuperscript{AP} method lines indicates the points of interest that we describe below.}
\label{bterctltxperday}
\end{figure}

As shown in Figure~\ref{bterctltxperday}, for the transaction frequency result, the T1 tainting results do not differentiate much from C1 results. Although the transaction frequency of FIFO\textsuperscript{AP}, LIFO\textsuperscript{AP} and TIHO\textsuperscript{AP} methods for T1 are still in the high percentile group as the majority of C1 cases have no more than 100 transactions per day.

All T1 tainting methods' results share very similar patterns throughout the whole tainting period and the number of transactions also stay considerably constant between 200 and 700 transactions for the period of interest. Although at one point FIFO\textsuperscript{AP} method results for T1 show the highest transaction frequency with 1,113 transactions on day 11 at the purple dot point as shown in Figure~\ref{bterctltxperday}, surpassing all other sample results.

Interestingly, there are two C1 cases that possess an extremely high number of transactions per day, as shown by the C1 high case one and two lines in Figure~\ref{bterctltxperday}. Both results show very different patterns when compared to each other and to the T1 tainting results. While C1 high case two shows extremely high transaction frequency on the first four-day period, the results show a gradual decrease until it reaches the majority of C1 cases on day 8. For C1 high case one results, the transaction frequency abruptly increases on day 6 but rapidly decrease after day 8. Nevertheless, it remains the sample case with highest transaction frequency until the end of tainting except for day 11.

\begin{figure}[h]
\centering
{
\resizebox*{7.5cm}{!}{\includegraphics{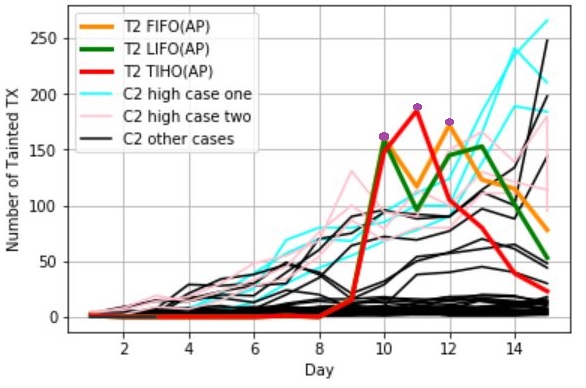}}}\hspace{5pt}
\caption{The transaction frequency per day in T2 and C2 tainting results. Each black line represents the results of each tainting method for each C2 case, and the cyan and pink line represent the C2 cases with the highest results which are referred as C2 high case one and two. The orange, green and red lines represent each tainting method results in T1. The three purple dots at day 10, 11 and 12 for T2 tainting methods' lines indicate the points of interest that we describe below.}
\label{betcoinctltxperday}
\end{figure}

As shown in Figure~\ref{betcoinctltxperday}, the results of the transaction frequency display some intriguing patterns for T2 tainting result. Interestingly, there are two C2 cases that also possess the highest number of transaction per day as shown by the C2 high case one and two lines. While both results show similar patterns when compare to other C2 cases during the first 12 day period. Both results suddenly exponentially increase on day 13 and become the samples with highest onward.

While there is no transaction activity for T2 during the first eight days, the transaction activity for all of the T2's FIFO\textsuperscript{AP}, LIFO\textsuperscript{AP} and TIHO\textsuperscript{AP} methods suddenly increase on day 9 and exceed all of the C2 results on day 10 at 150 transactions as marked at the first purple dot point as shown in Figure~\ref{betcoinctltxperday}. The number of tainted transactions for the T2's TIHO\textsuperscript{AP} method increases further and remains the highest on day 11, as shown at the second dot in Figure~\ref{betcoinctltxperday}, while the number of transactions in the other two tainting methods decreases, but nonetheless remain as the highest percentile sample. On day 12, the number of tainted transactions for the T2's FIFO\textsuperscript{AP} method become the highest, as shown at the third dot. Interestingly, the number of transactions of the T2's three tainting methods rapidly decrease afterwards and become more in line with the other C2 cases on day 14.

The sudden increase of the transaction frequency from zero in the first eight-day for the T2 results is behaviour that we did not expect. The explanation we have for this unexpected behaviour is that after waiting for about one year before starting to transfer the stolen Bitcoins, the thief waits for eight more days to check whether there are still tracking attempts or any interest from the public before proceeding further. Although, It is worth mentioning that the year 2013 is the time that the price of Bitcoin rapidly increases for the first time in its history from 10 US dollars in February to around 100 US dollars in April~\cite{bitcoinprice_2019}, Therefore, there is a possibility that the thief is merely waiting for the Bitcoin value to rise before deciding to spend the stolen coins.

%% MM: The following paragraph is not clear
Both C1 and C2 also have sample cases that show an exceedingly
high number of transactions. Our interpretation for the exceptionally high transaction frequency in the control cases is that due to the large transaction value criteria we used to select the control groups, sample the transactions that we select can belong to the service addresses, which would lead to two possibilities of what would happen subsequently as follows.

First, the service addresses transfer the Bitcoins to other addresses that also belong to the services, if the following addresses also match the criteria then the tainting would stop. Second, the service addresses distribute the Bitcoins to their users such as when the users exchange the real-world money for Bitcoins with cryptocurrency exchange services. This would result in widespread distribution of the tracked Bitcoins and exponentially increase the transaction frequency as the Bitcoins keep spreading as seen in the C2 high case one and two in Figure~\ref{betcoinctltxperday}, and the C1 high case one in Figure~\ref{bterctltxperday} to a degree.

Overall, while the transaction frequency variable results do not exactly match with H5 hypothesis that the sample theft cases would have a higher number of transactions per day than the control groups, and each tainting method results does not reveal the explicit distinction between the theft case samples and their control groups. The results, nonetheless, provide us with an interesting perspective on the theft transactions' unusual behaviours that can be used in future analysis.

\subsection{The Reused, Fresh and Service Addresses Metrics}

\begin{figure}[h]
\centering
{
\resizebox*{8cm}{!}{\includegraphics{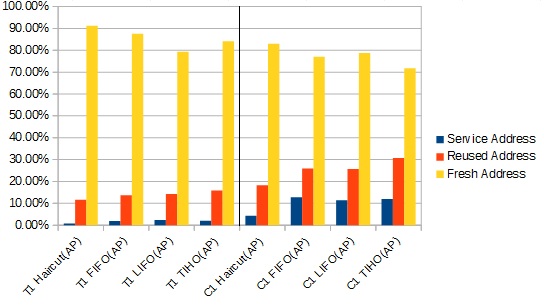}}}\hspace{5pt}
\caption{The address type percentage in T1 and C1 tainting results. The black line in the middle is the division line for the results of T1 and C1. Each bar represents the number of each type of the addresses(service address, reused address and fresh address) compared to the total number of tainted addresses in percentage. Poison\textsuperscript{AP} method result is merged into Haircut\textsuperscript{AP} method in the results.}
\label{bterctladdresstype}
\end{figure}

As shown in Figure~\ref{bterctladdresstype}, the percentage of the service, reused and fresh addresses are considerably different between T1 and C1 results. All of the T1 tainting methods results share a similar percentage for reused addresses at around 10\% and service addresses at one per cent. The reused address results of T1 tainting methods are also considerably lower than C2. The percentage of fresh addresses are relatively more significant in T1 except for the LIFO\textsuperscript{AP} method at 80\%, which is also similar to LIFO\textsuperscript{AP} method result for C1.

\begin{figure}[h]
\centering
{
\resizebox*{8cm}{!}{\includegraphics{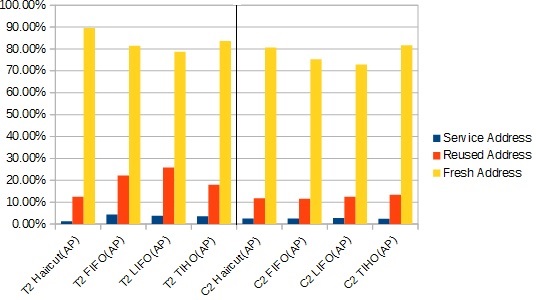}}}\hspace{5pt}
\caption{The address type percentage for T2 and C2 tainting results. The black line in the middle is the division line for the results of T2 and C2. Each bar represents the number for each type of the addresses compared to the total number of tainted addresses.}
\label{betcoinctladdresstype}
\end{figure}

%% MM: I do not understand what the 30% refer to, I see a maximum of 25% in Fig 9
As shown in Figure~\ref{betcoinctladdresstype}, the address type percentage for T2 tainting results reveal quite different patterns than for T1; the percentage of reused addresses for T2 is around 10\% higher than for C2 for all four tainting methods, with the LIFO\textsuperscript{AP} method being as high as 25\%. The fresh address percentage is also overall slightly higher for T2 results at around 80\%, while the C2 results are between 70\% and 80\%. The service address percentage is similar for T2 and C2, with the FIFO\textsuperscript{AP} method for T2 showing a slightly higher percentage at four per cent compared to the rest.

The reused variable results reveal conflicting yet interesting results for both T1 and T2 cases as shown in Figures~\ref{bterctladdresstype} and~\ref{betcoinctladdresstype} respectively. Both T1 and T2 display very distinctive results from their respective control groups, but in the opposite direction.

Our interpretation on the glaring contrast between the two cases' results is that it is possible for the stolen coins in T2 to pass through our service address profiling undetected and are already are in the possession of the other users. Another possibility is that the thief employs transaction obscuring technique that distributes the stolen coins in multiple looping patterns to make the analysis of the transaction network more difficult. In any case, the reused addresses results do not exactly match with H1 hypothesis that the sample theft cases would have a lower number of reused addresses than the control groups.

The fresh address variable results for both T1 and T2 cases considerably match with H2 hypotheses that the number of fresh addresses would be higher in theft sample cases than the control groups, as while all of the FIFO\textsuperscript{AP}, LIFO\textsuperscript{AP} and TIHO\textsuperscript{AP} methods show a higher number of fresh addresses in general compared to their control groups counterpart. However the difference between the two groups is not as significant as we expected, therefore, further analysis is required to ascertain whether the fresh address variable can present a definite contrast in theft transaction cases.

The service addresses variable results show a considerably high number for both T1 and T2 which match with H4 hypothesis that the number of service addresses in the sample theft cases would be high. There are two possible explanations for high service address number as follows; the service addresses that we detect may belong to the types of service directly involved in the transaction obscuring processes like laundering/tumbling services, CoinJoin services and coin mixing services. Another possibility is that the service addresses may be reused addresses of the common users with high enough transaction activity to reach the service address criteria.

Nevertheless, the results still prove there is already a certain number of tainted Bitcoins that manage to reach addresses that are likely to belong to service entities. This means that the integration of address profiling into the tainting method can improve the tainting method accuracy which provides an answer to our second research question of how can the tainting method be improved further for tracking of stolen Bitcoins. 

\subsection{The Number of Addresses per Transaction Metric}

\begin{figure}[h]
\centering
{
\resizebox*{7cm}{!}{\includegraphics{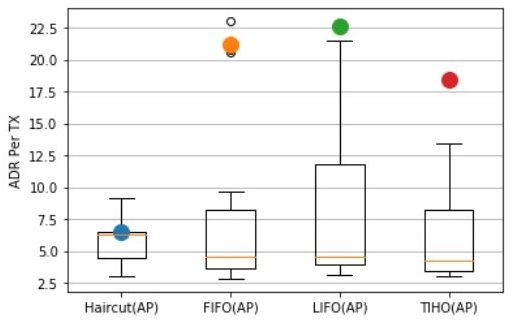}}}\hspace{5pt}
\caption{The number of addresses per transaction for T1 and C1 tainting results. Each box and its whiskers represent the data distribution of each C1 tainting method results. The white circles outside of the boxes are the outliers of C1 results, while the coloured circles represent the results of T1 tainting methods.}
\label{bterctladdresspertx}
\end{figure}

The FIFO\textsuperscript{AP}, LIFO\textsuperscript{AP} and TIHO\textsuperscript{AP} methods for T1 show remarkable results for addresses per transaction variable as shown in Figure~\ref{bterctladdresspertx}. All of the T1 tainting methods except the Haircut\textsuperscript{AP} method have a considerably higher number of addresses per transaction than the C1 cases at around 20 addresses per transaction. While there are also two other outliers in FIFO\textsuperscript{AP} for C1 that are close to the T1's FIFO\textsuperscript{AP} method results, the T1's LIFO\textsuperscript{AP} and TIHO\textsuperscript{AP} methods have the furthest outliers compared to the C1 results.

\begin{figure}[h]
{
\resizebox*{7cm}{!}{\includegraphics{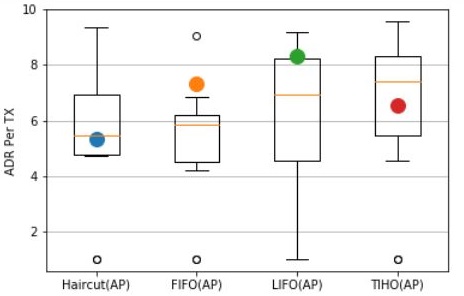}}}\hspace{5pt}
\caption{The number of addresses per transaction for T2 and C2 tainting results. Each box and its whiskers represent the distribution of each tainting method results for C2. The white circles outside of the boxes are the outliers for the C2 results, while the coloured circles represent the results for T2.}
\label{betcoinctladdresspertx}
\end{figure}

As shown in Figure~\ref{betcoinctladdresspertx}, for the addresses per transaction variable, the results do not reveal a distinction between the T2 and C2 as much as for the T1 case. The number of addresses per transaction for Haircut\textsuperscript{AP}, LIFO\textsuperscript{AP} and TIHO\textsuperscript{AP} tainting results for T2 are within a similar range as for the C2 tainting results. While FIFO\textsuperscript{AP} tainting results for T1 is the outlier of C2's FIFO tainting results, it is still not the furthest outlier compared to C2.

The addresses per transaction variable results also present a contradiction between T1 and T2 tainting results as shown in Figures~\ref{bterctladdresspertx} and~\ref{betcoinctladdresspertx} respectively. Our explanation on the contrast between T1 and T2 addresses per transaction variable results is that as the theft distribution transaction for T2 occurred in 2013, there are still not as many complex transaction obscuring methods available at the time that would allow the thief to disguise the transactions in the way as shown in T1 yet. For examples, the CoinJoin method is first purposed in by one of the Bitcoin developers in August 2013~\cite{gmaxwell_coinjoin_2013}.

As a result, while the number of addresses per transaction metric shown in this study yield promising results, the implementation of the number of addresses per transaction metric and H6 hypothesis still need to be revised and investigated further with a larger number of sample cases.

\subsection{The Transaction Fee Metric}

\begin{figure}[h]
\centering
{
\resizebox*{8cm}{!}{\includegraphics{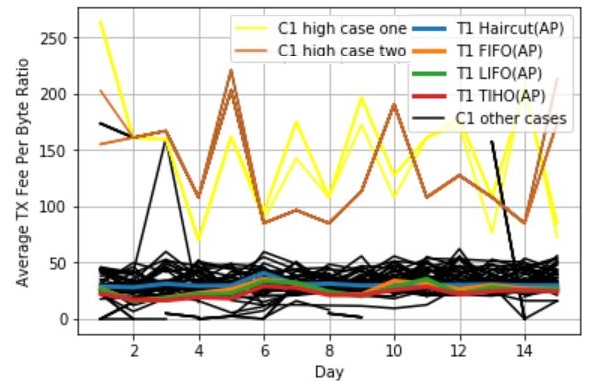}}}\hspace{5pt}
\caption{The average transaction fee per byte ratio for T1 and C1 tainting results.  Each black line represents the results of each tainting method for each C1 case, and the yellow and brown lines represent the C1 cases with the highest results which are referred as C1 high case one and two. The blue, orange, green and red lines represent each T1 tainting method result. It should be noted that the interested C1 cases presented in this result are not the same ones as presented in the C1 cases.}
\label{bterctltxfee}
\end{figure}

As shown in Figure~\ref{bterctltxfee}, T1 tainting methods results also do not appear to reveal distinction to C1 results for the transaction fee per byte ratio overall. All of the tainting methods for T1 also share a similar pattern and constantly stay in the lower percentile group with around 25 Sat per byte throughout the whole period of interest.

There are also two C1 cases that possess an extremely high average transaction fee per byte ratio in the C1 high case one and two lines as shown in Figure~\ref{bterctltxfee}. Both C1 high case one and two show very similar patter during the first six-day period but diverge later on but still stay within 100 - 200 Sat range until the end of our tainting. 

\begin{figure}[h]
\centering
{
\resizebox*{8cm}{!}{\includegraphics{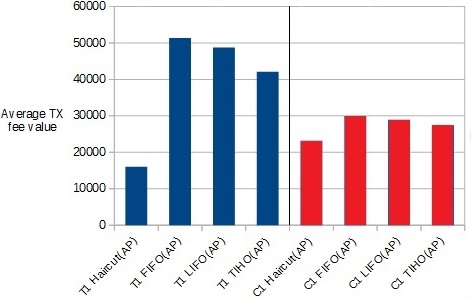}}}\hspace{5pt}
\caption{The average transaction fee value in T1 and C1 tainting results. The black line in the middle is the division line for the results of T1 and C1. The blue bars represent the results of T1 tainting, the red bars represent the results of C1 tainting.}
\label{bterctltxfeeval}
\end{figure}

As shown in Figure~\ref{bterctltxfeeval}, the average transaction fee value in the three tainting methods, FIFO\textsuperscript{AP}, LIFO\textsuperscript{AP} and TIHO\textsuperscript{AP} for T1, show significantly higher fee value than that of T1's Haircut\textsuperscript{AP} method and all of the C1 tainting methods. The average transaction fee value for T1's three tainting methods are in between 40,000 - 50,000 Sat range, while C1 tainting results are in between 20,000 - 30,000 Sat range.

\begin{figure}[h]
\centering
{
\resizebox*{8cm}{!}{\includegraphics{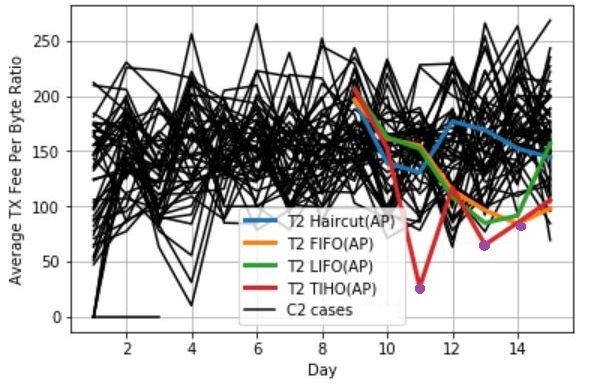}}}\hspace{5pt}
\caption{The average transaction fee per byte ratio for T2 and C2 tainting results. Each black line represents the results of each tainting method in each C2 case, while the blue, orange, green and red lines represent each T2 tainting method results. The three purple dots at day 11, 13 and 14 for T2 tainting methods lines indicate the points of interest that we discuss below.}
\label{betcoinctltxfee}
\end{figure}

As shown in Figure~\ref{betcoinctltxfee}, the transaction fee variable yields intriguing results for case T2. While the average transaction fee ratio of T2 tainting results are among the top percentile group when the transactions start appearing on day eight, the transaction fee ratio of the TIHO\textsuperscript{AP} method for T2 drops sharply and becomes the sample case with the lowest transaction fee at 25 Sat per byte in day 11 as pointed out at the purple dot in Figure~\ref{betcoinctltxfee}. The average transaction fee ratio for the TIHO\textsuperscript{AP} method increases on day 12 and then becomes the lowest on day 13 at the second purple dot. At the third purple dot on day 14, all of the three tainting methods for T2 have lower transaction fee values than for all of the C2 cases at around 80 Sat per byte.

\begin{figure}[h]
\centering
{
\resizebox*{8cm}{!}{\includegraphics{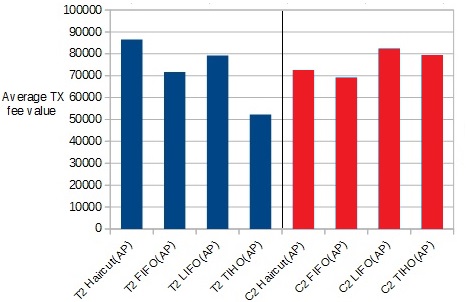}}}\hspace{5pt}
\caption{The average transaction fee value for T2 and C2 tainting results. The black line in the middle is the division line for the the results for T2 and C2. The blue bars represent the results of T2 tainting, the red bars represent the results of C2 tainting.}
\label{betcoinctltxfeeval}
\end{figure}

As shown in Figure~\ref{betcoinctltxfeeval}, although not significantly higher than its control groups as for T1, the average transaction fee value for T2 tainting methods are still as high as for C2 cases. Although, the TIHO\textsuperscript{AP} method for T1 still appear as the lowest of compared to the other T1's tainting methods and C1 cases. Both of T1 and C1 tainting methods have average transaction fee value between 70,000 - 80,000 Sat range, while the average fee value for T1's TIHO\textsuperscript{AP} method is at 52,000 Sat.

In general, the transaction fee per size ratio of both T1 and T2 tainting methods are in the lower percentile group when compare to their respective control groups as shown in Figures~\ref{bterctltxfee} and~\ref{betcoinctltxfee}. Although the results between T1 and T2 are quite different in that the transaction fee ratio results of all T1 tainting methods are remarkably constant among other the lower percentile C1 cases, while T2 shows unstable changes for all of the tainting methods throughout the tainting period especially for TIHO\textsuperscript{AP}.

However, despite the overall low transaction fee ratio for both T1 and T2 which means that both theft cases transactions pay the fee at the lower rate than the majority of their respective control groups, the transaction fee in both cases are still in the higher percentile as shown in Figures~\ref{bterctltxfeeval} and~\ref{betcoinctltxfeeval}.

Our explanation for this difference is that the reason why the transaction fee per size ratio is considerably lower in T1 is due to the fact that the majority of transaction in T1 are considerably larger than the C1 transactions as can be seen from the number of addresses per transaction variable results even though the average transaction fee value in T1 is higher than C1. This interpretation can also be applied for T2 as the number of addresses per transaction in T2 FIFO\textsuperscript{AP} and LIFO\textsuperscript{AP} methods are also in the higher percentile, while TIHO\textsuperscript{AP} is in the lower percentile at the first quantile.

While the transaction fee variable show very promising potential and can discover clear unusual behaviours in the theft transactions, the transaction fee results still do not completely match with H3 hypothesis that the sample theft cases would have higher transaction fee value than the control groups. Additionally, all of the T1 tainting methods' results still yield very similar results in the average transaction fee ratio, despite the significant number of non-overlapping tainted transactions in T1 which we still cannot yet provide a conclusive explanation for this similarity. Therefore, a further extensive investigation into the transaction fee with broader sample cases is still required before we can truly evaluate the performance of this variable.

\subsection{The Overlapping Tainted Transactions}

\begin{figure}[h]
\centering
{
\resizebox*{7cm}{!}{\includegraphics{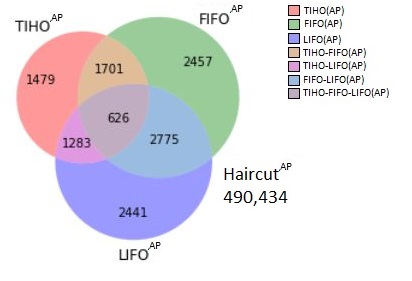}}}\hspace{5pt}
\caption{The number of overlapping tainted transactions between each T1 tainting method. Each circle represents the results of the tainted transaction to each tainting method and the overlapping area represent the tainted transaction results that the tainting methods share. For example, the red circle represent the number of tainted transactions results of TIHO\textsuperscript{AP} method, and the green circle represent the results of  FIFO\textsuperscript{AP} method. The yellow overlap area between the two circles represents the number of transactions that both methods classify as tainted.}
\label{overlaptrabter}
\end{figure}

As shown in Figure~\ref{overlaptrabter}, all of the tainting methods possess balance portions between overlapping and non-overlapping tainted transaction results for the T1 case. The tainted transaction results of TIHO\textsuperscript{AP} method are slightly more similar to the FIFO\textsuperscript{AP} method than to the LIFO\textsuperscript{AP} method.

\begin{figure}[h]
\centering
{
\resizebox*{6.5cm}{!}{\includegraphics{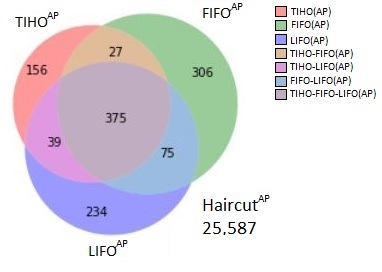}}}\hspace{5pt}
\caption{The number of overlapping tainted transactions between each of the T2 tainting methods. Each circle represents the result of the tainted transaction of each tainting method and the overlapping area represent the tainted transaction results that the tainting methods share.}
\label{overlaptrabetcoin}
\end{figure}

As shown in Figure~\ref{overlaptrabetcoin}, the overall pattern of overlapping transactions' results for T2 is relatively similar to that of T1. Although the portions of overlapping between each tainting methods are more significant in this case as the the overlapping number between the three tainting methods is larger than each method's non-overlapping number. Each tainting method still has a considerable portion of tainted transactions that is unshared by the other methods. Interestingly, the FIFO\textsuperscript{AP} and LIFO\textsuperscript{AP} methods share more similarity in the tainted transactions' results than with the TIHO\textsuperscript{AP} method.

Overall, the three tainting methods FIFO\textsuperscript{AP}, LIFO\textsuperscript{AP} and TIHO\textsuperscript{AP} demonstrate the significantly diverse results of tainting methods for both T1 and T2 cases as shown in Figures~\ref{overlaptrabter} and~\ref{overlaptrabetcoin}. This implies that there is a significant number of mixing involve between the tainted and clean Bitcoins for both sample cases. Ultimately, the tracking results of each tainting method would be completely different in the longer tainting period. The overall tainting results of T1 and T2 are shown in Tables~\ref{Btertaintresult} and~\ref{Betcointaintresult} in the appendix.

% \subsection{Tainting Methods Comparison}

% Hence, we present a comparison summary for the performance of each tainting method based on the hypothesis that we presented as follows; The result of The FIFO\textsuperscript{AP} method results show a higher transaction frequency in both T1 and T2 cases which according to H5 hypothesis would mean that the FIFO\textsuperscript{AP} perform better than the other three in this aspect. FIFO\textsuperscript{AP} method also show higher number of fresh addresses for both case T1 and T2 cases. Interestingly, the TIHO\textsuperscript{AP} method is mostly the method with lowest number in most variable such as the transaction frequency, transaction fee and number of addresses per transactions for both T1 and T2

\section{Conclusion and Future Work} \label{sec:conclusion}

While the privacy that Bitcoin can bring to users is revolutionary in today’s modern society, the privacy features can also facilitate the individuals to perform illegal activities or even cause harm to others. In an attempt to combat crime and illegal activities in Bitcoin, tracing the Bitcoins to the end of the blockchain alone would only show who are the unfortunates to be the last holders of tainted Bitcoins chosen by the tainting process. To truly track the crimes in Bitcoin, it is crucial to understand the context of each transaction and address involved.

The variables we use for tainting and evaluation in this paper are information that can be found within the blockchain. Such information cannot be falsified due to the nature of the blockchain. Although it is possible to obtain additional information from the external sources that are available in public such as forums or social media websites~\cite{fleder_bitcoin_2015}, there is a risk of the information being either incorrect or purposely falsified, so extra caution must be exercised when handling information from external sources.

The result of our experiment shows that some of the evaluation variable results are in conflict with our hypotheses and there are some contrasts between the results of the two sample theft cases that we study. The evaluation metrics and the hypotheses that we proposed still require further analysis and validation before we can truly measure and evaluate the tracking accuracy of the tainting methods. Additionally, the number of sample theft cases can be expanded further to improve the evaluation results. Nevertheless, the evaluation metrics demonstrate the potential for the future application of illegal activities tracking in cryptocurrency.

Another consideration that should be taken into account is that illegal activities are one of the most important aspects of the Bitcoin economy, considering that as high as 33\% of all Bitcoin transactions involve illegal activities~\cite{Foley_sex_2018}. Thus, the classification of service addresses as an exit point of tainted transactions that utilises only transaction activity may not be sufficient enough, considering the thieves would more likely prefer spending the stolen coins on the exit points with the least chance of being caught, as opposed to official services like cryptocurrency exchange services which could be regulated by  governments. As the services or businesses that engage in illegal activities are likely to employ transaction obscuring techniques to protect their users' privacy, the address profiling should be developed further by incorporating other techniques such as address clustering and network analysis to assist in the address profiling and tainting process. The network analysis result of the tainted transaction network can also be used to discern the flow of complex transactions and the relationship of the addresses that are concealed by transaction obscuring techniques~\cite{baumann2014exploring}.

This paper laid the foundation for our future work on taint analysis to not only discover the most accurate tainting methods but to also improve upon the current transaction tracking strategies. The results of transaction taint analysis can then be used to assist the cybersecurity in combating against cryptocurrency cybercrimes. This will have important implications not only to cybersecurity but to financial regulatory developments.

\bibliographystyle{IEEEtran}
%\bibliography{IEEEabrv,mybibliography}


\newpage
\appendices
\section*{Appendix}
\begin{table}[h]
\centering
\scriptsize
 \caption{The transaction hash of the control transaction for each theft case.}
\begin{tabular}{|l|}
\hline
\textbf{Bter Theft Case TX Hash (T1)}                             \\ \hline
e297b80d69c824d91c8d5a05b4026e38ebcd3b5502730ece078f1d51ff1d455c \\ \hline          \textbf{Control TX Hash (C1)}                                              \\ \hline
f530aff14bfd6e0b0ff7c12fff870382edc175c1eabd604f1b4776d3360dce0c \\ \hline
f6cc6c3041a1928153f6c678eeeb2ad62a09edc8bcb957f9ec58f77016b6dc95 \\ \hline
6f96299e9afb900a8fb1c319d4c9b7bfce23cd2485cabf81702e66cc59635d26 \\ \hline
5731123a378f630cab213f8619dbb6af362c45039cce2dcf153ca459edf86327 \\ \hline
938a9d776f80fe0361d4b75b658b5605548eb8b42286f1eed9c74dd1e0b30d7c \\ \hline
0e8a6bb0abc54ccf6cd771344e61b3f6f79261a08a8465dfb900bbd4c3bb793c \\ \hline
5b839ef2255c1ecc1d0cf0b63ea04e0b626a926da1d6ed5d9c430a07f02635e3 \\ \hline
3295935d13f1836522df42e90c07be74f4d254a11d730a0d18d8c2d022c6c3d7 \\ \hline
b2421e5660a233d8f00333acc7fb1699c48fcae4ea7ce72f9d825a45a89985d7 \\ \hline
e297b80d69c824d91c8d5a05b4026e38ebcd3b5502730ece078f1d51ff1d455c \\ \hline
b735489d580befe81f5b40e6b3cac29f273a3d22236843db0c963ba06e28dc4e \\ \hline
0e0ed5de5a456aa8bed99200033bf8e03bd8f68244e0020f34f39d6bdab82fac \\ \hline
6b3da870bb890abdaa3756006e73529c60c3d995c45ae172b9ee0182f9d4d674 \\ \hline
e1e73d158b5fd3c93e6d65014d769501344deb2fbc0055189f5c895877af44f4 \\ \hline
79bc949bf4b2d63fc29cbb44690f548c247d7baf187be3b79a4ab13fa635517e \\ \hline
8496b6302b9f68153a18db80c646614ffcbc60ac4fb0c3af345dd27007df59dd \\ \hline
929ef91f1a189ff28614d56036b043291ae2fb4769ae16a348a38d3c1cc79806 \\ \hline
f51569462ac213245672ba0e0ba95977ab9c98fbf1e6a7edc455171638d93499 \\ \hline
8ceeeb0f3b7bf6de77efb293d80c011dee5a5a15cc0acf3542be12a507472726 \\ \hline
f1ced97f8e988be90a032aeef7e57afc94beca9a0f98808ce696899f14577f2e \\ \hline
34f4dd143074589768b8206eedf1825b83373aa36142d1a60239dd334e56c6af \\ \hline
034870270941ea2cc9b8ea610378ae5558e27560457da1f3ac3d59216e2ca02e \\ \hline
f95309dde7d89c55415afc49fd55705d426ee6ce0b0a7e6a9e02cb02d50afcc6 \\ \hline
9a3c39145d5ed6d352a860fd36a89bab9d4237ef2cbe1b90f6d6ea178fa0d1e2 \\ \hline
1b8e6a420408dca6704b1ac37cbcc653d4c814d6e24d76cd4be8f8d9806df7b5 \\ \hline
\textbf{Betcoin theft Case TX Hash (T2)}                          \\ \hline
21ac34ced0f55701b38a3f8e02b9d2b8451db2f0071cd5c761a857d1ac0a78df \\ \hline
\textbf{Control TX Hash (C2)}                                              \\ \hline
36d6e70e5d9128d341c445bf83e5057dbba70145ba2bdf56fc4b0d246cb89c8c \\ \hline
d27e267faa01b6a720fb679430ae88bbbeaded5f1c17f67d1282bcc926303a4f \\ \hline
32570bdb6ab20bd11fc77486007e4cd76d23a2ebee2e87b4ce5117bddf8586bc \\ \hline
63810ef189609de75716d5001ecc0fc08d0581156c1515ff01bbf425ba20b707 \\ \hline
8e67511f1434a8ba61fbd96da071fde8cd208ffef4fa54fdcde8f68e0164ef81 \\ \hline
62de8da526b7a0d58009be8543d31f0e3fa1386267c8557d9615040351945773 \\ \hline
de063e761d93b1d47c013365032b4ea40076940f76923d431751b4918603f573 \\ \hline
a6a44979af76435f1200ea1a34aa664ab03a67260a218a3c975b5c933cb962f5 \\ \hline
9b450bd0637593a292836da0675ab3abed841a4df62fd2171f0b76fa459df2d3 \\ \hline
737826368507f3674d11b9b804828807b2a53c33a51ea27d1ded0de040facafc \\ \hline
4f577677974c4ad44358de124a6e5bb76f83921be040591b0dfd07eb83de1b90 \\ \hline
f067e8073f26d56d806733e5500e8339d62d75a65b86b99263be0080332d4e5e \\ \hline
b3561f02f212ebca98dfe3134094435f87a8f387698c43fb2bb002cccc8d4984 \\ \hline
6d2b97f11cc4a074c26b1a588ed6a8047b8b68445cc8d03a7cf31961dceb8a45 \\ \hline
\end{tabular}
\label{txhashtable}
\end{table}

Table~\ref{txhashtable} shows the complete list of transaction hash of each sample theft case and control sample that we select in this paper for each theft sample case.

\begin{table}[h]
\caption{The results of each tainting method on case T1.}
\begin{center}
\begin{tabular}{|m{1.52cm}|c|c|c|c|}
\hline
\textbf{\textit{Variables}} &\textbf{\textit{Haircut\textsuperscript{AP}}} & \textbf{\textit{FIFO\textsuperscript{AP}}} & \textbf{\textit{LIFO\textsuperscript{AP}}} & \textbf{\textit{TIHO\textsuperscript{AP}}} \\
\hline
Tainted TX & 502,196      &7,559  & 7,125 & 5,089\\
\hline
Tainted ADR  & 1,128,107    &13,606    & 11,685 &9,058 \\
\hline
Service ADR   & 9,136     &261    & 278 & 183 \\
\hline
Reused ADR  & 132,084   &1,869  & 1,670 &1,434 \\
\hline
Fresh ADR  & 102,8627   &11,918  & 9,269 &7,622 \\
\hline
Avg. ADR Per TX  & 6.49   &21.17  & 22.62 & 18.45 \\
\hline
Avg. TX Fee Value(Sat) & 16,038.22   &51,383.05  & 48,761.31 & 42,109.70\\
\hline
\end{tabular}
\label{Btertaintresult}
\end{center}
\end{table}

Table~\ref{Btertaintresult} shows the overall result of each T1 tainting method from 15 days tainting.

\begin{table}[h]
\caption{The results of each tainting method in case T2}
\begin{center}
\begin{tabular}{|m{1.65cm}|c|c|c|c|}
\hline
\textbf{\textit{Variables}} &\textbf{\textit{Haircut\textsuperscript{AP}}} & \textbf{\textit{FIFO\textsuperscript{AP}}} & \textbf{\textit{LIFO\textsuperscript{AP}}} & \textbf{\textit{TIHO\textsuperscript{AP}}} \\
\hline
Tainted TX   & 26,799      &783  & 723 & 597\\
\hline
Tainted ADR    & 44,245    &815    & 755 &619 \\
\hline
Service ADR   & 564    &36    & 29 & 22 \\
\hline
Reused ADR  & 5,509   &181  & 195 &111 \\
\hline
Fresh ADR  & 39,612   &664  & 594 &518 \\
\hline
Avg. ADR Per TX  & 5.33   &7.32 & 8.34 & 6.50 \\
\hline
Avg. TX Fee Value(Sat)  & 86,542.66  & 71,685.82 & 79,183.95 & 52,311.56 \\
\hline
\end{tabular}
\label{Betcointaintresult}
\end{center}
\end{table}

Table~\ref{Betcointaintresult} shows the overall result of each T2 tainting method from 15 days tainting.

\end{document}